# Propagator Of A Particle In A Photonic And A Magnetic Field And Evolution Of A Coherent State

E.G. THRAPSANIOTIS

Department of Applied Mathematics
University of Crete,
71409 Heraklion, Crete, GREECE.

E-mail: egthra@hotmail.com

**ABSTRACT**
In the present paper we study the interaction of a photonic field with a particle in a magnetic field. We use formalism similar to the Born-Oppenheimer approximation of molecular physics and we eliminate the photonic field variables by using a method related to the Berry phase to obtain an effective Hamiltonian and then extract the propagator of the particle. As an initial photonic state we assume a coherent one.

## 1. Introduction

In recent years Berry phase [1-6], quantum invariant forms [7-12] and propagators [13-18] have attracted the interest of many researchers in various areas of physics. In the present paper we combine the above areas of research in the study of the interaction of a single mode field with a particle in the presence of a magnetic field.

We adopt a similar scheme as in [19-21] for the interaction of radiation with a particle. We proceed by eliminating the field variables for a specific initial field state, considering the initial particle+field system uncoupled and expanding the field on an over complete coherent basis. The whole approach is based on the fact that coherent states are stable states of the whole part of the Hamiltonian involving field terms. Then we use standard methods [22-23] to extract the propagator of the particle in the photonic field.

The paper proceeds in the following order. In section 2 we study the invariant form and the stable states of the terms of the Hamiltonian involving field variables. In section 3 we extract the effective Hamiltonian and the propagator of the particle. Finally in section 4 we give our conclusions.

## 2. Time dependent quantum invariant forms and stable states.

The time dependent evolution of coherent states has been studied extensively in the past years. Generally it is interesting if an initial state remains stable under its time evolution. At least instinctively we can assume that a stable coherent state is related with a quantum invariant form of the Hamiltonian that describes the time evolution of the coherent state referred above. This topic is discussed in literature [9] and we give some of the main results in this section, as they are vital for the subsequent discussion.

In fact for a given Hamiltonian $H(t)$ a hermitian invariant form $I(t)$ is defined as

$$\frac{dI}{dt} = \frac{\partial I}{\partial t} + \frac{1}{i}[I,H] = 0 \qquad (1)$$

with $I^+ = I$. Suppose that the operator $I$ exists and has eigenstates $|\lambda,t\,)$ with eigenvalues $\lambda$. Then if $I$ do not involve differentiation then it is possible to select a phase so that the eigenstates

$$|\lambda,t\rangle = e^{i\Phi_\lambda(t)}|\lambda,t\,) \qquad (2)$$

satisfy the time dependent Schrodinger equation. Additionally the phase satisfies the equation

$$\frac{d\Phi_\lambda}{dt} = \left(\lambda,t\,\left|i\frac{\partial}{\partial t} - H\right|\lambda,t\,\right) \qquad (3).$$

The term $\int_0^t \left(\lambda,t'\,\left|i\frac{\partial}{\partial t'}\right|\lambda,t'\,\right)dt'$ corresponds to the Berry phase.

Let us consider the Hamiltonian

$$H_0(t) = \omega a^+ a + g(t)a + g^*(t)a^+ \qquad (4)$$

and suppose that the quantum invariant form $I$ can be written as the following bilinear form

$$I(t) = a^+ a - \vartheta(t)a^+ - \vartheta^*(t)a + \chi(t) \qquad (5).$$

On replacing this form in (1) we find that the following equations must be valid

$$i\dot\vartheta = \omega\vartheta - g^* \qquad (6a)$$

$$i\dot\chi = \vartheta^* g - \vartheta g^* = i\frac{d|\vartheta|^2}{dt} \qquad (6b).$$

Now we define the operators

$$b = a - \vartheta(t) \qquad b^+ = a^+ - \vartheta^*(t) \qquad (7a)$$

$$[b,b^+] = [a,a^+] = 1 \qquad (7b)$$

so that $I$ can be written as

$$I = b^+ b + (\chi - |\vartheta|^2) \qquad (8)$$

Consequently the eigenstates of $I$ are the same with those of the operator $b^+ b$, so that

$$I|m,t\,) = \lambda_m |m,t\,) \qquad \lambda_m = m + (\chi - |\vartheta|^2) \qquad (9)$$

where the state $|m,t\,)$ is defined as
$$b^+ b|m,t\,) = m|m,t\,) \qquad (10a)$$
$$b|m,t\,) = \sqrt{m}|m-1,t\,) \qquad b^+|m,t\,) = \sqrt{m+1}|m+1,t\,) \qquad (10b).$$
We can easily prove that the state $|0,t\,) = |\vartheta(t))$ is a coherent state, where according to (6a)
$$\vartheta(t) = \vartheta(0)e^{-i\omega t} - ie^{-i\omega t}\int_0^t g^*(t')e^{i\omega t'}dt' \qquad (11).$$
Additionally the time dependent Schrodinger equation of the Hamiltonian (4) is obeyed by
$$|\vartheta(t)\rangle = e^{i\Phi_0(t)}|\vartheta(t)) \qquad (12a)$$
where
$$\Phi_0(t) = -\int_0^t \mathrm{Re}(g(t')\vartheta(t'))dt' \qquad (12b).$$

### 3. Effective hamiltonian and propagator of the particle.

In the model of the interaction of radiation with matter studied in this section we apply formalism similar to that of the molecular Born-Oppenheimer theory.

The Hamiltonian can be written as a sum of three terms. The particle Hamiltonian $H_p$, the single mode field one $H_f$, and the interaction term $H_I$
$$H = H_p + H_f + H_I \qquad (13).$$
Particularly the particle Hamiltonian is given as
$$H_p = \frac{1}{2m}\left(\vec{p} - \frac{1}{c}\vec{A}\right)^2 \qquad (14a)$$
where
$$\vec{A} = \frac{1}{2}\vec{H} \times \vec{r} \qquad (14b)$$
and $\vec{H} = (0,0,H)$ is the magnetic field vector in the z-direction.
The Hamiltonian of the radiation field has the form
$$H_f = \omega a^+ a \qquad (15)$$
and finally, the interaction Hamiltonian is given as
$$H_I = -e\vec{r}\cdot\vec{E}_f \qquad (16).$$
The second quantized form of the field operator is given as
$$\vec{E}_f(\vec{r}) = \frac{1}{\sqrt{V}}il(\omega)\hat{\varepsilon}\left[ae^{i\vec{k}\cdot\vec{r}} - a^+e^{-i\vec{k}\cdot\vec{r}}\right] \qquad (17)$$
where $V$ is the quantization volume and $l(\omega)$ is a real function of frequency given as $l(\omega) = \sqrt{\hbar\omega/2\varepsilon_0}$. $\hat{\varepsilon}$ is the polarization along the z-axis. In the present case we adopt the dipole (long wavelength) approximation ($e^{i\vec{k}\cdot\vec{r}} \approx 1$ in eq. 17), so that the field operator has the form

$$\vec{E}_f = \frac{1}{\sqrt{V}} i l(\omega)\hat{\varepsilon}(a - a^+) \qquad (18)$$

and $H_I$ takes the form

$$H_I = -\frac{1}{\sqrt{V}} i e l(\omega)\hat{\varepsilon}\cdot\vec{r}(t)(a - a^+) \qquad (19).$$

Now we combine the terms (15) and (19) involving field variables in the term

$$H_0(a^+, a; t) = H_f + H_I = \omega a^+ a + g(t)a + g^*(t)a^+ \qquad (20a)$$

where

$$g(t) = -\frac{1}{\sqrt{V}} i e l(\omega)\hat{\varepsilon}\cdot\vec{r}(t) \qquad (20b).$$

The total Hilbert space can be written as the direct product of the particle and field parts

$$H = H^{particle} \otimes H^{field}$$

From the commutation relations

$$[\hat{r}, a] = [\hat{r}, a^+] = 0 \qquad (21a)$$

$$[\hat{r}, H_0] = 0 \qquad (21b)$$

we conclude that we can have the basis sets

$$|\vec{r}, \alpha\rangle \qquad (22a)$$

$$|\vec{r}, n\rangle \qquad (22b)$$

where

$$\hat{r}|\vec{r}, n\rangle = \vec{r}|\vec{r}, n\rangle \qquad (23a)$$

$$H_0(\hat{r})|\vec{r}, n\rangle = \varepsilon_n(\vec{r}, t)|\vec{r}, n\rangle \qquad (23b).$$

$|\vec{r}, \alpha\rangle$ are eigenvectors of the operator $a$ as well as the position. Here $n$ represents the field quantum numbers that describe the energy $\varepsilon_n(\vec{r}, t)$. We can expand $|\vec{r}, n\rangle$ with respect the basis $|\vec{r}, \alpha\rangle$

$$|\vec{r}, n\rangle = \int d^3r' \int d^2\alpha' |\vec{r}', \alpha'\rangle\langle\vec{r}', \alpha'|\vec{r}, n\rangle \qquad (24a)$$

$$\langle\vec{r}', \alpha'|\vec{r}, n\rangle = \delta(\vec{r} - \vec{r}')\langle\alpha'|n\rangle_{\vec{r}} \qquad (24b).$$

Let $|\psi^E\rangle$ be the total wave function which satisfies the Schrodinger equation

$$H|\psi^E\rangle = E|\psi^E\rangle \qquad (25).$$

Then

$$\langle\vec{r}, n|H_p + H_0|\psi^E\rangle = E\langle\vec{r}, n|\psi^E\rangle = E\psi_n^E(\vec{r}, t) \qquad (26)$$

$$\langle\vec{r}, n|H_0|\psi^E\rangle = \varepsilon_n(\vec{r}, t)\psi_n^E(\vec{r}, t) \qquad (27).$$

In the direct product space the operator $H_p$ can be written as

$$H_p = H_p^{particle} \otimes \hat{1}^{field}$$

the basis vectors of which are
$|\vec{r},n\rangle = |\vec{r}\rangle \otimes |n;\vec{r}\rangle$.

Here $\{|\vec{r}\rangle\}$ represents a basis in the space $H^{particle}$ while $\{|n;\vec{r}\rangle\}$ represents a basis in the space $H^{field}$, the elements of which are eigenvalues of the $r$ dependent $H_0$ that acts on the space $H^{field}$

$H_0|n;\vec{r}\rangle = \varepsilon_n(\vec{r},t)|n;\vec{r}\rangle$  (28).

On using these expressions we have the following calculations

$$\langle \vec{r},n|H_p \otimes \hat{1}|\psi^E\rangle =$$
$$= \langle \vec{r},n|H_p \otimes \sum_m \int d^3r'|\vec{r}',m\rangle\langle\vec{r}',m|\psi^E\rangle =$$
$$= \langle n;\vec{r}|\sum_m \int d^3r'|m;\vec{r}'\rangle\langle\vec{r}|\frac{1}{2m}\left(\hat{p}-\frac{1}{c}\vec{A}\right)^2|\vec{r}'\rangle\langle\vec{r}',m|\psi^E\rangle =$$
$$= \langle n;\vec{r}|\sum_m \int d^3r'|m;\vec{r}'\rangle\left[\langle\vec{r}|\frac{1}{2m}\left(-i\nabla_{\vec{r}}-\frac{1}{c}\vec{A}\right)^2|\vec{r}'\rangle\right]\langle\vec{r}',m|\psi^E\rangle =$$
$$= \langle n;\vec{r}|\sum_m \left[\frac{1}{2m}\left(-i\nabla_{\vec{r}}-\frac{1}{c}\vec{A}\right)^2\right]\int d^3r'|m;\vec{r}'\rangle\delta(\vec{r}-\vec{r}')\langle\vec{r}',m|\psi^E\rangle =$$
$$= \sum_m \langle n;\vec{r}|\left[\frac{1}{2m}\left(-i\nabla_{\vec{r}}-\frac{1}{c}\vec{A}\right)^2\right]|m;\vec{r}'\rangle\langle\vec{r}',m|\psi^E\rangle$$

(29)

where the differential operator $\nabla_{\vec{r}}$ acts on both terms on the right. Now we define the non-Abelian vector potential [6]

$\vec{A}_{mn}(\vec{r}) = i\langle n;\vec{r}|(\nabla_{\vec{r}}|m;\vec{r}\rangle)$   (30)

and the covariant momentum

$\hat{\pi}_{mn} = \delta_{mn}\left(\hat{P}-\frac{1}{c}\vec{A}\right) - \vec{A}_{mn}$   (31a)

which in the representation of position, of the particle variables can be written as

$\vec{\pi}_{mn} = \delta_{mn}\left(-i\nabla_{\vec{r}}-\frac{1}{c}\vec{A}\right) - \vec{A}_{mn}$   (31b).

Additionally we have

$$\langle m;\vec{r}|\left(-i\nabla_{\vec{r}}-\frac{1}{c}\vec{A}\right)^2|n;\vec{r}\rangle = \sum_l \left(\delta_{ml}\left(-i\nabla_{\vec{r}}-\frac{1}{c}\vec{A}\right) - \vec{A}_{ml}\right) \cdot \left(\delta_{ln}\left(-i\nabla_{\vec{r}}-\frac{1}{c}\vec{A}\right) - \vec{A}_{ln}\right) =$$
$$= \sum_l \vec{\pi}_{ml} \cdot \vec{\pi}_{ln}$$

(32).

Now we suppose that the initial field state is a coherent state, so that

$$|\psi(0)\rangle = |\vartheta(0)\rangle\psi_0(\vec{r}) \quad (33)$$

where $\psi_0(\vec{r})$ is the initial particle wave function.

We proceed by using the Born – Oppenheimer trick and consider as the "fast" part the coherent photonic field, corresponding to the electrons in the conventional Born – Oppenheimer approximation. Then we expand the state vector $|\psi(t)\rangle$ on the eigenstates of the Hamiltonian $H_0$ involving the photonic part. The ground state of those eigenstates corresponds to a coherent state and as in the present paper we suppose that the initial photonic state is a coherent one the whole problem reduces to the question if the photonic field is going to remain in such a state as the system evolves. We observe that the Hamiltonian $H_0$ does not commute with the Hamiltonian $H_p$ so that the stability of the photonic field in a coherent state cannot be exact. By correspondence with the conventional Born – Oppenheimer theory we expect that the above mentioned stability is relevant to the topology of the energies $\varepsilon_n(\vec{r},t)$ in (28) as $\vec{r}$ varies and more particularly to the question if $\varepsilon_0(\vec{r},t)$ has any touch or cross with other bands. This problem has attracted many discussions. If the value of $\vartheta(0)$ and value of the volume are large enough we should expect to there be a region of space where no touch or cross of $\varepsilon_0(\vec{r},t)$ with other bands appear. Then in the conventional Born – Oppenheimer theory the stability is relevant with the mass of the atoms in the molecule. The square root of the inverse of that is a measure of how stable the energies will be in a region free of any touch or cross. More details can be found in ref. [24] and in the references there. Certainly we should expect that a very large volume would make the coherent state almost completely stable as the interaction energy in (20) would become extremely small. In that case the stability time would depend on the velocity of the particle as well.

Proceeding further to the present problem and if the above discussion is valid then we can expect that (33) can be set to evolve to the state

$$|\psi(t)\rangle = |\vartheta(t)\rangle e^{i\Phi_0(t)}\psi_0(\vec{r},t) = |\vartheta(t)\rangle\psi_0(\vec{r},t) \quad (34)$$

where according to (11)

$$\vartheta(t) = \vartheta(0)e^{-i\omega t} + \frac{1}{\sqrt{V}}el(\omega)e^{-i\omega t}\int_0^t dt' e^{i\omega t'}\hat{\varepsilon}\cdot\vec{r}(t') \quad (35a)$$

$$\Phi_0(t) = -\int_0^t dt' \operatorname{Re}(g(t')\vartheta(t')) \quad (35b).$$

The time evolution of an arbitrary field state can be obtained by expanding on a set of coherent states as

$$|\chi(0)\rangle = \int \frac{d^2\alpha}{\pi} \langle\alpha|\chi(0)\rangle|\alpha\rangle$$

$$|\chi(t)\rangle = \int \frac{d^2\alpha}{\pi} \langle\alpha|\chi(0)\rangle|\eta(t)\rangle e^{i\Phi_0(t)} \quad (36)$$

where

$$\eta(t) = \alpha e^{-i\omega t} + S(\omega)e^{-i\omega t}\int_0^t dt' e^{i\omega t'}\hat{\varepsilon}\cdot\vec{r}(t') \quad (37a)$$

$$\Phi_0(t) = S(\omega)\int_0^t dt'\,\text{Re}\big(i\alpha e^{-i\omega t'}\hat{\varepsilon}\cdot\vec{r}(t')\big) + S^2(\omega)\int_0^t dt'\int_0^{t'} ds\,\text{Re}\big(ie^{-i\omega(t'-s)}\hat{\varepsilon}\cdot\vec{r}(t')\hat{\varepsilon}\cdot\vec{r}(s)\big)$$

$$(37b)$$

$$S(\omega) = \frac{1}{\sqrt{V}} el(\omega) \quad (37c).$$

Now we extract the equation that $\psi_0(\vec{r},t)$ obeys. On replacing (34) in the time dependent Schrödinger equation we intend to extract an adiabatic effective [6] Hamiltonian describing the motion of the particle in the photonic field. The magnetic field is added naturally as an external field. Then we are going to have the following equations

$$i\frac{\partial|\psi(t)\rangle}{\partial t} = i\frac{\partial|\vartheta(t)\rangle}{\partial t}\psi_0(\vec{r},t) + i\frac{\partial\psi_0(\vec{r},t)}{\partial t}|\vartheta(t)\rangle = (H_p + H_0)|\vartheta(t)\rangle\psi_0(\vec{r},t) \quad (38),$$

$$i\frac{\partial|\vartheta(t)\rangle}{\partial t} = H_0|\vartheta(t)\rangle \quad (39).$$

So upon simplifying (38) by using (39) and then taking from the left the projection on the bra $\langle\vartheta(t)| = \langle 0;\vec{r}|$ and using (32), we conclude that we have the following sequence of equations

$$i\frac{\partial\psi_0(\vec{r},t)}{\partial t} = \left[\frac{1}{2m}\sum_l \vec{\pi}_{0l}\cdot\vec{\pi}_{l0}\right]\psi_0(\vec{r},t) \quad (40a)$$

$$i\frac{\partial\psi_0(\vec{r},t)}{\partial t} = \left[\frac{1}{2m}\left(i\nabla_{\vec{r}} + \frac{1}{c}\vec{A} + \vec{A}_{00}\right)^2 + \frac{1}{2m}\sum_{l\neq 0}\vec{A}_{0l}\cdot\vec{A}_{l0}\right]\psi_0(\vec{r},t) \quad (40b)$$

$$i\frac{\partial\psi_0(\vec{r},t)}{\partial t} = \left[\frac{1}{2m}\left(i\nabla_{\vec{r}} + \frac{1}{c}\vec{A} + \vec{A}_{00}\right)^2 + \frac{1}{2m}\left(\begin{array}{c}-\vec{A}_{00}\cdot\vec{A}_{00} + \\ +(\nabla_{\vec{r}}\langle\vartheta(t)|)\cdot(\nabla_{\vec{r}}|\vartheta(t)\rangle)\end{array}\right)\right]\psi_0(\vec{r},t)$$

$$(40c)$$

where we have used the fact that
$$\vec{A}_{00} = iS(\omega)\hat{\varepsilon}\big(\vartheta(t) - \vartheta^*(t)\big) \quad (41).$$

In deriving (41) we have taken into account the definition of $|\vartheta(t)\rangle$ and (30).

Then on using relations (12a) and (41), we can further extract from (40c) the Hamiltonian describing the time evolution of $\psi_0(\vec{r},t)$. It is

$$H' = \frac{1}{2m}\vec{p}^2 - i\frac{S(\omega)}{2m}[\vartheta(t) - \vartheta^*(t)]\hat{\varepsilon}\cdot\vec{p} - i\frac{S(\omega)}{2m}\hat{\varepsilon}\cdot\vec{p}[\vartheta(t) - \vartheta^*(t)] +$$
$$+ \frac{S^2(\omega)}{m}\left[\frac{1}{2} + |\vartheta(t)|^2 - \frac{1}{2}\vartheta^{*2}(t) - \frac{1}{2}\vartheta^2(t)\right] - \frac{1}{2mc}(\vec{A}\cdot\vec{p} + \vec{p}\cdot\vec{A}) + \quad (42a)$$
$$+ \frac{1}{2mc^2}\vec{A}^2 + \frac{1}{mc}\vec{A}\cdot\vec{A}_{00}$$

where

$$\vartheta(t) = \vartheta(0)e^{-i\omega t} + S(\omega)e^{-i\omega t}\int_0^t dt' e^{i\omega t'}\hat{\varepsilon}\cdot\vec{r}(t') \quad (42b).$$

If the magnetic field is on the z-axis
$$\vec{A} = \frac{1}{2}\hat{\varepsilon}_y Hx - \frac{1}{2}\hat{\varepsilon}_x Hy \quad (43)$$
where $\hat{\varepsilon}_x$ and $\hat{\varepsilon}_y$ are respectively on the x and y axis. $\hat{\varepsilon}$ is on the z-axis as noted above. Consequently in this case we have two decoupled problems and the Hamiltonian can be written as
$$H' = H_0^{x,y} + H_0^z \quad (44)$$
where

$$H_0^{x,y} = \frac{p_x^2 + p_y^2}{2m} + \frac{H}{2mc}(yp_x - xp_y) + \frac{H^2}{8mc^2}(x^2 + y^2) \quad (45a)$$

$$H_0^z = \frac{p_z^2}{2m} - i\frac{S(\omega)}{2m}[\vartheta(t) - \vartheta^*(t)]\hat{\varepsilon}\cdot\vec{p} - i\frac{S(\omega)}{2m}\hat{\varepsilon}\cdot\vec{p}[\vartheta(t) - \vartheta^*(t)] +$$
$$+ \frac{S^2(\omega)}{m}\left[\frac{1}{2} + |\vartheta(t)|^2 - \frac{1}{2}\vartheta^{*2}(t) - \frac{1}{2}\vartheta^2(t)\right] \quad (45b)$$

with the propagator of $H_0^{x,y}$ given as

$$K_{x,y}(x_f, y_f; x_i, y_i; t) = \frac{m\omega_1}{4\pi i \sin(\omega_1 t/2)}\exp\left\{\begin{array}{l}\frac{im\omega_1}{4}\cot\left(\frac{\omega_1 t}{2}\right)[(x_f - x_i)^2 + (y_f - y_i)^2] + \\ + i\frac{m\omega_1}{2}(x_i y_f - x_f y_i)\end{array}\right\}$$
(46a)

where
$$\omega_1 = \frac{H}{c} \quad (46b).$$

Now we proceed to a certain approximation of the exact till now theory to take a final expression for $H_0^z$ and its propagator as well. On performing the change of variables $\tau = t - t'$ we obtain the following integral in (42b)

$$e^{-i\omega t}\int_0^t dt' e^{i\omega t'}\hat{\varepsilon}\cdot\vec{r}(t') = \int_0^t d\tau e^{-i\omega\tau}\hat{\varepsilon}\cdot\vec{r}(t-\tau) \quad (47)$$

Now we perform the Taylor expansion
$$\hat{\varepsilon}\cdot\vec{r}(t-\tau) = \hat{\varepsilon}\cdot\vec{r}(t) - \tau\hat{\varepsilon}\cdot\dot{\vec{r}}(t) + \ldots \quad (48)$$

Then after substitution of the Taylor expansion in the integral (47) the first term $\vec{r}(t)$ can be taken out of the integral and we obtain the expression

$$\vartheta(t) = \vartheta(0)e^{-i\omega t} + 2\frac{S(\omega)}{\omega}e^{-i\omega t/2}\sin\left(\frac{\omega t}{2}\right)\vec{\varepsilon}\cdot\vec{r}(t) \qquad (49)$$

This is equivalent to the approximation used in [20]. Higher order terms in the Taylor expansion are in fact negligible as they are going to involve powers of $S(\omega) = \frac{1}{\sqrt{V}}el(\omega)$, and consequently of the volume, and their contribution is going to be of order $S^2(\omega)$ and on, as the last term in (42b) already involve an $S(\omega)$ contribution. This fact can be proved by taking into account the equation of motion that can be derived from the decoupled Hamiltonian (44) via standard methods. Consequently after certain manipulations we obtain

$$H_0^z = \frac{p_z^2}{2m} + \psi(t)(zp_z + p_z z) + \frac{1}{2}m\omega^2(t)z^2 + e(t)p_z + f(t)z + h(t) \qquad (50)$$

where

$$\psi(t) = -2\frac{S^2(\omega)}{m\omega}\sin^2\left(\frac{\omega t}{2}\right) \qquad (51a)$$

$$\omega^2(t) = \frac{16S^4(\omega)}{m^2\omega^2}\sin^4\left(\frac{\omega t}{2}\right) \qquad (51b)$$

$$e(t) = -i\frac{S(\omega)}{m}\left[\vartheta(0)e^{-i\omega t} - \vartheta^*(0)e^{i\omega t}\right] \qquad (51c)$$

$$f(t) = i\frac{S^3(\omega)}{m\omega}\sin^2\left(\frac{\omega t}{2}\right)\left[\vartheta(0)e^{-i\omega t} - \vartheta^*(0)e^{i\omega t}\right] \qquad (51d)$$

$$h(t) = \frac{S^2(\omega)}{m}\left[\frac{1}{2} + |\vartheta(0)|^2 - \frac{1}{2}\vartheta^2(0)e^{-2i\omega t} - \frac{1}{2}\vartheta^{*2}(0)e^{2i\omega t}\right] \qquad (51e)$$

(50) has the following propagator [22,25]

$$K_z(z_f,t;z_i,0) = \sqrt{\frac{i}{2\pi c_3 c_2}}\exp\left\{\frac{(z_i/c_2 + d_1 - z_f)^2}{2ic_3} + \frac{ic_1}{2}z_i^2 + id_2 z_f + id_3\right\} \qquad (52)$$

where [22,25]

$$c_1(t) = \frac{\partial}{\partial t}(\ln F(t)), \quad c_1(0) = 0 \qquad (53a)$$

$$c_2(t) = \exp\left[2\int_0^t \psi(s)ds\right] \times \left|\frac{F(t)}{F(0)}\right| \qquad (53b)$$

$$c_3(t) = -\int_0^t d\rho \exp\left[-4\int_0^\rho \psi(s)ds\right] - F^2(0)\int_0^t \frac{d\rho}{F^2(\rho)} \qquad (53c)$$

and $F(t)$ is given by the equation [25]

$$F''(t) + 4\psi(t)F'(t) + \omega^2(t)F(t) = 0 \qquad (54)$$

In the present case

$$F(t) = A\exp\left(\frac{4\pi}{mV}t - \frac{2\pi}{m\omega V}\sin(2\omega t)\right)\left(MathieuC(0,q,z) + FFMathieuS(0,q,z)\right)$$
(55a)

where

$$q = -\frac{2\pi}{m\omega V} \quad (55b)$$

$$z = \frac{\omega t}{2} - \frac{\pi}{4} \quad (55c)$$

$$FF = \frac{MathieuC\,Prime\left(0,q,\frac{\pi}{4}\right)}{MathieuS\,Prime\left(0,q,\frac{\pi}{4}\right)} \quad (55d)$$

and A is a constant.
We use the symbols of Mathematica [26] for the Mathieu functions that appear here, as in the present case we have an initial value problem. Finally we obtain the following expressions for c and d in (52)

$$c_1(t) = \frac{4\pi}{mV} - \frac{4\pi}{mV}\cos(2\omega t) + \\ + \frac{\omega}{2}\frac{MathieuC\,Prime(0,q,z) + FFMathieuS\,Prime(0,q,z)}{MathieuC(0,q,z) + FFMathieuS(0,q,z)}$$
(56a)

$$c_2(t) = \left|\frac{MathieuC(0,q,z) + FFMathieuS(0,q,z)}{MathieuC\left(0,q,\frac{\pi}{4}\right) - FFMathieuS\left(0,q,\frac{\pi}{4}\right)}\right| \quad (56b)$$

and

$$c_3(t') = -\int_0^{t'}\exp\left[\frac{8\pi}{mV}t - \frac{4\pi}{m\omega V}\sin(2\omega t)\right]dt - \\ -\int_0^{t'}\exp\left[-\frac{8\pi}{mV}t + \frac{4\pi}{m\omega V}\sin(2\omega t)\right]\frac{\left(MathieuC\left(0,q,\frac{\pi}{4}\right) - FFMathieuS\left(0,q,\frac{\pi}{4}\right)\right)^2}{\left(MathieuC(0,q,z) + FFMathieuS(0,q,z)\right)^2}dt$$
(56c)

$$d_1(t) = \frac{1}{i}\int_0^t b_1(u)du \quad (57a)$$

$$d_2(t) = \frac{1}{i}\int_0^t b_2(u)du \quad (57b)$$

$$d_3(t) = -\frac{1}{i}\int_0^t b_2(u)d_1(u)du + \frac{1}{i}\int_0^t b_3(u)du \quad (57c)$$

where

$$b_1(t) = -ie(t)\frac{1}{c_2(t)} + if(t)c_3(t)c_2(t) \quad (58a)$$

$$b_2(t) = -ie(t)c_1(t)\frac{1}{c_2(t)} - if(t)c_2(t) \quad (58b)$$

$$b_3(t) = -ih(t) \quad (58c).$$

The above expressions describe the interaction of radiation with a particle in a magnetic field. Upon using the fact that we can obtain the evolution of $\psi_0(\vec{r},t)$ from the product of propagators (46) and (52) we conclude that these equations and (34) can give the wave function of the particle studied at any time.

## 4. Conclusions

In the present paper we studied a model Hamiltonian that describes the interaction of radiation with a particle in a magnetic field, and we extracted the corresponding propagator in a closed form, so that we can find the time evolution of an initially prepared wave packet.

The original point in the present work is the use of stable coherent states in the part of the Hamiltonian involving field terms to manipulate the field variables.

Such a method is not directly applicable in the case of the quadratic field Hamiltonian studied previously by the author in [19,21] via path integral methods, as the differential equations involved there are not elementarily solvable [7]. We intend to study this case by using the present formalism numerically in a subsequent paper.

Additionally we notice that the results of the present paper are similar with the ones that we derived in [20]. There we used path integral methods while in the present paper we use methods based on the geometric phase. In the present paper we add a magnetic field as well.

We think that the present model is simple and tractable and gives new aspects of photonic processes.